\documentclass[10pt]{revtex4}
\usepackage{amssymb}
\usepackage{latexsym}
\usepackage{epsfig}                                   
\usepackage{amsmath}

\begin{document}

\title{Horizon Thermodynamics and Gravitational Field Equations in
Quasi-Topological Gravity}
\author{A. Sheykhi $^{1,2}$\thanks{
asheykhi@shirazu.ac.ir}, M. H. Dehghani$^{1,2}$ \thanks{
mhd@shirazu.ac.ir} and R. Dehghani $^{1}$}
\address{$^1$  Physics Department and Biruni Observatory, College of
Sciences, Shiraz University, Shiraz 71454, Iran\\
         $^2$  Research Institute for Astronomy and Astrophysics of Maragha
         (RIAAM), P.O. Box 55134-441, Maragha, Iran}
\begin{abstract}
In this paper we show that the gravitational field equations of $(n+1)$%
-dimensional topological black holes with constant horizon
curvature, in cubic and quartic quasi-topological gravity, can be
recast in the form of the first law of thermodynamics,
$dE=TdS-PdV$, at the black hole horizon. This procedure leads to
extract an expression for the horizon entropy as well as the
energy (mass) in terms of the horizon radius, which coincide
exactly with those obtained in quasi-topological gravity by
solving the field equations and using the Wald's method. We also
argue that this approach is powerful and can be extended to all
higher order quasi-topological gravity for extracting the
corresponding entropy and energy in terms of horizon radius.\\

Keywords: quasi-topological; thermodynamics; gravity.
reconstruction.
\end{abstract}
\maketitle

\section{Introduction\label{Intr}}

In recent years, most theoretical physicists as well as
cosmologists, have been convinced that there should be a deep
relation between the gravitational field equations and the laws of
thermodynamics. It was pointed out by Jacobson  for the first time
in $1995$, that the hyperbolic second order partial differential
Einstein equation has a predisposition to the first law of
thermodynamics \cite{Jac}. Indeed, Jacobson derived the Einstein
field equations of general relativity, in its tensorial form, by
applying the Clausius relation, $\delta Q=T\delta S$,
on the horizon of spacetime, here $\delta S$ is the change in the entropy and $%
\delta Q$ and $T$ are the energy flux across the horizon and the
Unruh temperature seen by an accelerating observer just inside the
horizon. Also, by applying the Clauius relation to the apparent
horizon of the Friedmann-Robertson-Walker universe, the
corresponding Friedmann equations can be derived in Einstein,
Gauss-Bonnet and more general Lovelock gravity \cite{CaiKim}.
Following these investigations, a lot of attempts have been done
to reveal the connection between thermodynamics and gravity in
different setups
\cite{Par,Pad,Cai0,Cai1,Cai2,Cai3,Cai4,Shey1,Shey2,ShHL}. For
example, in Ref. \cite{CaiHL} the relationship between the first
law of thermodynamics and the gravitational field equation of a
static, spherically symmetric black hole in Horava-Lifshitz
gravity has been explored. It was shown that, the gravitational
field equations of static, spherically symmetric black holes in
Horava-Lifshitz theory can be expressed as the first law of
thermodynamics on the event horizon \cite{CaiHL}. This approach
can lead to extract expressions for the entropy and mass of
Horava-Lifshitz black holes which are consistent with those
obtained from other approaches \cite{CaiHL}. These results further
support the idea that gravitation on a macroscopic scale is a
manifestation of thermodynamics.

It is well known that the natural generalization of the
Einstein-Hilbert action to higher dimensional spacetime, and
higher order gravity with second order equation of motion, is the
Lovelock action \cite{Lov}. However, because of the topological
origin of the Lovelock terms, the second term of the Lovelock
action (the Gauss-Bonnet term) does not have any dynamical effect
in four dimensions. Similarly, the cubic interaction only
contributes to the equations of motion when the bulk dimension is
seven or greater. Recently, a modification of third order Lovelock
gravity was proposed by \cite{Olive1,Myer1} which contains cubic
terms of Riemann tensor and contribute to the equation of motions
in five dimensions. This new theory, which is called
``quasi-topological" gravity, was also extended to include the
quartic terms of Riemann tensor \cite{MHD2}. Quasi-topological
gravity provides a useful toy model to study a broader class of
four (and higher)
dimensional CFT's, involving three or more independent parameters \cite%
{Myer2}. Black hole spacetimes in higher order quasi-topological gravity
which have at most second order derivatives of the metric in the field
equations, have been explored and their thermodynamics have been
investigated \cite{MHD1,MHD3,MHD4,QT}.

In this paper we turn to investigate the connection between the
gravitational field equations and the first law of thermodynamics in
quasi-topological gravity. For a static topological black hole spacetime
with constant horizon curvature, we will show that the gravitational field
equations can be transformed to the first law of thermodynamics, $TdS-dE=PdV$%
, on the black hole horizon. This allows us to extract the entropy
expression in terms of the horizon radius, which is useful in
studying thermodynamics of these kind of black holes.

The structure of this paper is as follow. In the next section, we
show that the gravitational field equations in Einstein and
Gauss-Bonnet gravity can be recast as the first law of
thermodynamics on the black hole horizon. In section III, we
briefly review the action of the quasi-topological gravity. In
section IV, we apply the method to cubic quasi-topological theory.
In section V, we will generalize our approach to the quartic and
higher order quasi-topological gravity and extract the entropy and
energy expressions of these theories. We finish our paper with
conclusions in section VI.

\section{Horizon thermodynamics in Einstein and Gauss-Bonnet
gravity} Let us start with Einstein-Hilbert and Gauss-Bonnet cases
to set the stage and to see how the method works \cite{Par}. One
can derive the equations of motion for gravitational theory by
either, varying the action of the theory with respect to $g_{\mu
\nu}$, without specifying the form of the spacetime metric, or by
specifying the spacetime metric, then inserting the metric in the
action, and finally varying the resulting action with respect to
the metric functions. In both cases one arrives at the field
equations. To see how the two approaches leads to the same result,
in this section, we use both the field equations as well as
variation method, for transforming the equations of motion to the
first law of thermodynamics at the black hole horizon.
\subsection{The Einstein-Hilbert Case}
The Einstein field equations is ($c=1$)
\begin{equation}\label{eins22}
R_{ab}-\frac{1}{2} R g_{ab}=8 \pi G T_{ab}.
\end{equation}
We consider a four-dimensional static, spherically symmetric
spacetime with a horizon, which is described by the metric
\begin{equation}\label{met2}
ds^{2}=-f(r)dt^{2}+\frac{dr^{2}}{f(r)}+r^{2}d\Omega^{2}.
\end{equation}%
Inserting metric (\ref{met2}) in Eq. (\ref{eins22}), the $(rr)$
component of the Einstein equations can be written
 \begin{equation}\label{eins23}
\frac{1}{r^2}\left[r f^{\prime}(r) -1+f(r)\right]=8 \pi G P,
\end{equation}
where $P=T^r_{r}$ is the radial pressure of matter at the horizon
\cite{CaiHL}. Here prime denotes derivative with respect to $r$.
We assume that the spacetime has a  horizon at $r = r_{+}$ which
is the simple root of $f(r_{+})=0$. We also propose at $r = r_{+}$
the surface gravity $\kappa= f^{\prime}(r_{+})/2$ has non zero
value which implies a finite non zero temperature
$T={f^{\prime}(r_{+})}/{4\pi}$ at the horizon. Evaluating  the
field equation (\ref{eins23}) at the horizon where $f(r_{+})=0$,
we find
\begin{equation}\label{eins24}
\frac{1}{G}\left(\frac{r_{+} f^{\prime}(r_{+})}{2}
-\frac{1}{2}\right)=4\pi r_{+}^2 P .
\end{equation}
Multiplying both sides of Eq. (\ref{eins24}) by $dr_{+}$, we
arrive at
\begin{equation}\label{eins25}
\underbrace{\frac{f^{\prime}(r_{+})}{4 \pi}}_\text{T}
\underbrace{\frac{1}{G}d\left(\frac{4 \pi
r_{+}^2}{4}\right)}_\text{dS}
\underbrace{-\frac{1}{2}\left(\frac{dr_{+}}{G}\right)}_\text{-dE}=
\underbrace{P d\left(\frac{4\pi r_{+}^3}{3}\right)}_\text{P d V}.
\end{equation}
If we invoke the expressions for the entropy and energy (mass) of
the black hole,
\begin{eqnarray}\label{ent26}
S&=&\frac{A}{4G}, \\
E&=&\frac{r_{+}}{2G}, \label{enrg27}
\end{eqnarray}
where $A=4\pi r_{+}^2$ is the area of the horizon, we find that
Eq. (\ref{eins25}) is nothing but the first law of thermodynamics,
\begin{eqnarray}\label{Einfirstlaw}
dE=TdS-PdV.
\end{eqnarray}
\subsection{The Gauss-Bonnet Gravity}
We shall now turn our attention to the more general case, namely
the Gauss-Bonnet gravity. In an effective action approach to the
string theory, the Gauss-Bonnet term corresponds to the leading
order quantum corrections to gravity, and its presence guarantees
a ghost-free action\cite{zwiebach}. This theory contains a special
combination of curvature-squared term, added to the
Einstein-Hilbert action. The Gauss-Bonnet term does not have any
dynamical effect in four dimensions since it is just a topological
term in four dimensions. Static black hole solutions of
Gauss-Bonnet gravity have been found and their thermodynamics have
been investigated in ample details \cite{Caigb}. The action of the
Einstein- Hilbert in the presence of the Gauss-Bonnet correction
term, in $(n+1)$-dimensions, is given by
\begin{equation}\label{actg1}
I=\frac{1}{16\pi G_{n+1}}\int
{d^{n+1}x\sqrt{-g}\left(-2\Lambda+R+\alpha\mathcal{L}_{GB}
\right)}+\int d^{n+1}x\mathcal{L}_{M},
\end{equation}%
where $\alpha $ is the Gauss-Bonnet coefficient with dimension $($length$%
)^{2} $, and $\mathcal{L}_{GB}$ is the Gauss-Bonnet Lagrangian
which has the form,
\begin{equation}\label{actg1}
\mathcal{L}_{GB}=R_{abcd}R^{abcd}-4R_{ab}R^{ab}+R^{2}.
\end{equation}%
The field equations can be obtained by varying the above action
with respect to the metric $g_{ab}$. We find
\begin{equation}\label{gbfeq}
G_{ab}+\Lambda g_{ab}+2\alpha H_{ab}=8 \pi G_{n+1} T_{ab},
\end{equation}
where,
\begin{eqnarray}\label{eqeins}
&&G_{ab}=R_{ab}-\frac{1}{2} R g_{ab},\\
&&H_{ab}=RR_{ab}-2R_a{}^cR_{bc}-2R^{cd}R_{acbd}+R_a{}^{cde}R_{bcde}-\textstyle{1\over4}g_{ab}{\cal
L}_{GB},\label{eqgb1}
\end{eqnarray}
are the Einstein and the second-order Lovelock tensor,
respectively. Consider again a static spherically symmetric
solution of the form
\begin{equation}\label{met3}
ds^{2}=-f(r)dt^{2}+\frac{dr^{2}}{f(r)}+r^{2}d\Omega_{n-1}^{2}.
\end{equation}%
Substituting metric (\ref{met3}) in Eq. (\ref{gbfeq}), the $(rr)$
component of the field equations reduces to
\begin{equation}\label{eqmo2}
\frac{1}{2}\frac{(n-1)}{r^2}
\Big{\{}rf^{\prime}-\left(n-2\right)\left(1-f\right)+\frac{\tilde{\alpha}}{r^2}\left(1-f\right)\left[2r
f^{\prime}-\left(n-4\right)\left(1-f\right)\right]-\frac{nr^2}{l^2}\Big{\}}=8
\pi G_{n+1} P.
\end{equation}
where again $P=T^r_{r}$ is the radial pressure of matter at the
horizon, and we have defined $\tilde{\alpha}=(n-2)(n-3)\alpha$.

On the other hand, by substituting the metric (\ref{met3}) in
action (\ref{actg1}), and varying the action (\ref{actg1}) with
respect to $g^{rr}$, after multiplying both sides in
$(-g)^{-1/2}g^{rr}$, we get
\begin{equation}\label{Pressure2}
\frac{(n-1)}{16\pi G_{n+1}}\frac{1}{r^{2}}\Bigg{\{}\left( r+\frac{2\tilde{%
\alpha}}{r}-\frac{2\tilde{\alpha}f}{r}\right) f^{\prime }-\left(
n-2\right) \left(1-f\right) -\frac{\tilde{\alpha}}{r^2}\left(
n-4\right) (1-f)^2-\frac{nr^2}{l^2} \Bigg{\}}%
=P.
\end{equation}%
where $P$ is defined as,
\begin{equation}\label{P}
 P=T_{r}^{r}=g^{rr}T_{rr}=g^{rr}\left\{\frac{2}{\sqrt{-g}}\frac{\delta
\mathcal{L}_{M}}{\delta g^{rr}}\right\}.
\end{equation}%
As one can see, the resulting equation derived by variational
principle in (\ref{Pressure2}) is \textit{precisely} the same as
one obtained in (\ref{eqmo2}) directly from the field equations
(\ref{gbfeq}). As we mentioned already, we have two approaches for
deriving the components of the field equations; varying the action
with respect to $g_{\mu \nu}$, and then inserting metric in the
field equations, or by substituting the metric in the action, and
varying the resulting action with respect to the metric functions.

The variational method leads to (\ref{Pressure2}), can also be
applied to the more general metric
\begin{equation}\label{met1}
ds^{2}=-N^{2}(r)f(r)dt^{2}+\frac{dr^{2}}{f(r)}+r^{2}d\Sigma
_{k,n-1}^{2},
\end{equation}%
where $d\Sigma _{k,n-1}^{2}$ represents the line elements of an
$(n-1)$-dimensional constant curvature hypersurface with unit
radius. Without loss of the generality, one may take $k=1,-1,$ and
$0$, corresponding to spherical, hyperbolic and planar
hypersurface. Inserting metric (\ref{met1}) in action
(\ref{actg1}), after variation with respect to $g^{rr}$, we arrive
at
 \begin{eqnarray}\label{Pressure3}
&&\frac{(n-1)}{16\pi G_{n+1}}\frac{1}{Nr^{3}}\Bigg{\{}\left( r^{2}+2k\tilde{%
\alpha}-2\tilde{\alpha}f\right) \left( N^{2}f\right) ^{\prime
}+N^{2}\left[ r\left( n-2\right) \left( f-k\right)
-\frac{\tilde{\alpha}}{r}\left( n-4\right)
(k^{2}-2kf+f^{2})-\frac{nr^{3}}{l^{2}}\right] \Bigg{\}}  \notag \\
&&=T_{r}^{r}=P,
\end{eqnarray}%
where again the radial pressure is given by (\ref{P}). We also
assume the function $f(r)$ has a simple zero at $r=r_{+}$ with
$f(r=r_{+})=0$ and non-vanishing surface gravity $\kappa
=Nf^{\prime }(r_{+})/2$. The temperature
associated with the horizon is now defined as $T={\kappa }/{2\pi }={Nf^{\prime }(r_{+})}/{%
4\pi }$. Evaluating Eq. (\ref{Pressure3}) at $r=r_{+}$, we obtain
\begin{equation}\label{var022}
\frac{N(n-1)}{16\pi G_{n+1}r_{+}^{2}}\left[ f^{\prime }(r_{+})\left( r_{+}+%
\tilde{\alpha}\frac{2k}{r_{+}}\right) -k\left( n-2\right) -\frac{\tilde{%
\alpha}}{r_{+}^{2}}k^{2}\left( n-4\right)
-\frac{nr_{+}^{2}}{l^{2}}\right] =P.
\end{equation}
Consider two equilibrium states of the system with an
infinitesimal different in the extensive variables entropy,
energy, and volume $dS$, $dE$ and $dV$, respectively, while the
values of the intensive quantities of the system are the
temperature $T$ and pressure $P$. Our aim is to introduce in Eq.
(\ref{var022}) a factor $dV$ and see whether we can rewrite it in
the form $TdS-dE=PdV$. Multiplying both sides of Eq.
(\ref{var022}) by the factor $\Sigma _{k}r_{+}^{n-3}dr_{+}$, where
$\Sigma _{k}$ is the area of a unit $(n-1)$-dimensional constant
hypersurface with volume $V=\Sigma _{k}r_{+}^{n}/n$, we arrive at
\begin{equation}\label{var011}
\frac{\kappa }{2\pi }d\left[ \frac{\Sigma _{k}r_{+}^{n-1}}{4G_{n+1}}\left( 1+%
\frac{(n-1)}{(n-3)}\frac{2k\tilde{\alpha}}{r_{+}^{2}}\right)
\right] -d\left[
\frac{(n-1)\Sigma _{k}r_{+}^{n-2}}{16\pi G_{n+1}}\left( k+\frac{\tilde{\alpha%
}k^{2}}{r_{+}^{2}}+\frac{r_{+}^{2}}{l^{2}}\right) \right] =P\Sigma
_{k}r_{+}^{n-1}dr_{+}=PdV.
\end{equation}%
The first term in the left hand side is in the form $TdS$ and our
analysis allows us to read off the expression of entropy $S$ for
the horizon as,
\begin{eqnarray}\label{flt07}
S&=&\frac{\Sigma _{k}r_{+}^{n-1}}{4G_{n+1}}\left( 1+\frac{(n-1)}{(n-3)}\frac{2%
\tilde{\alpha}k}{r_{+}^{2}}\right).
\end{eqnarray}
In addition, the second term in (\ref{var011}) can be interpreted
as $dE$, where the energy of the system is given by
\begin{eqnarray}
E&=&\frac{(n-1)\Sigma _{k}r_{+}^{n-2}}{16\pi G_{n+1}}\left( k+\frac{\tilde{%
\alpha}k^{2}}{r_{+}^{2}}+\frac{r_{+}^{2}}{l^{2}}\right).
\label{flt08}
\end{eqnarray}%
Thus, we have transformed the field equations in Guass-Bonnet
gravity to the first law of thermodynamics, $TdS-dE=PdV$, on the
black hole horizon. The obtained expressions for the entropy and
energy in (\ref{flt07}) and (\ref{flt08}), coincide with the
expressions of entropy and energy for Gauss-Bonnet black holes in
AdS spaces derived by solving the field equations \cite{Caigb}.

In our analysis, we have supposed that both $l^2=-n(n-1)/{\Lambda}$
and $\tilde{\alpha}$ are fixed. Recently, there were a lot of interest
in denoting $\Lambda$ as a thermodynamical variable proportional to the pressure \cite{PVcrit}.
In this case the last term in the left-hand side of Eq. (\ref{var011}), $n(n-1)/(16\pi G_{n+1}l^2)dV$, may be
moved to the right-hand side as $P_{\Lambda}dV$. One should note that
the authors of Ref. \cite{PVcrit} have considered the solutions of Einstein field equations,
and therefore only the term $P_{\Lambda}dV$ will be appeared in the right-hand side of Eq. (\ref{var011}) for the
solutions of Einstein equation. Also, one may note that $P_{\Lambda}$ is the $T^r_r$-component of energy momentum tensor
of cosmological constant term, if one denotes the $\Lambda$-term as energy term in the right-hand side of Einstein equation.
Also, if one denotes $\tilde{\alpha}$ as a thermodynamical variable, then a term $\tilde{\alpha}dA$
will appear in the right hand side of Eq. (\ref{var011}), where $A$ is the conjugate quantity
to the Gauss-Bonnet coefficient \cite{CaiPV}. In this case, since the Gauss-Bonnet coefficient has
not the dimensions of pressure in geometric units, its conjugate quantity has not the dimensions of volume.

Here, before going to the case of quasi-topological gravity, we
pause to give a comment on the expressions for energy density
$\rho $ and pressure $P$. In general, these quantities are not the
same for a general perfect fluid. Indeed $P$ is given in Eq.
(\ref{Pressure3}), while $\rho $ can be calculated by varying the
action (\ref{actg1}) with respect to $g_{tt}$ and
multiplying both sides in $(-g)^{-1/2}g_{tt}$ as%
\begin{equation}\label{rho}
\frac{(n-1)}{16\pi G_{n+1}}\frac{N}{r^{3}}\Bigg{\{}\left( r^{2}+2k\tilde{%
\alpha}-2\tilde{\alpha}f\right) f^{\prime }+ r\left( n-2\right)
\left( f-k\right) -\frac{\tilde{\alpha}}{r}\left( n-4\right)
\left[
k^{2}-2kf+f^{2}\right] -\frac{nr^{3}}{l^{2}} \Bigg{\}}%
=-T_{t}^{t}=\rho.
\end{equation}%
Of course, one may note that the pressure and energy density are
the same on the horizon as one may see by calculating $\rho $ on
the horizon.
\section{Quasi-Topological Gravity\label{Quasi1}}
The action of the quasi-topological theory in $(n+1)$-dimensions
is given by
\begin{equation}\label{Act1}
I=\int d^{n+1}x\left( \mathcal{L}_{G}+\mathcal{L}_{M}\right) ,
\end{equation}%
where $\mathcal{L}_{M}$ is the Lagrangian of the matter and
\begin{equation}\label{act02}
\mathcal{L}_{G}=\frac{\sqrt{-g}}{16\pi G_{n+1}}\left( -2\Lambda +{\mu }_{1}%
\mathcal{L}_{1}+{\mu }_{2}\mathcal{L}_{2}+{\mu }_{3}\mathcal{X}_{3}+{\mu }%
_{4}\mathcal{X}_{4}+...\right) .
\end{equation}%
In the above equation $\Lambda =-(n-2)(n-3)/2l^{2}$ is the
cosmological
constant, $\mathcal{L}_{1}=R$ is the Einstein-Hilbert Lagrangian, $\mathcal{L%
}_{2}=R_{abcd}{R}^{abcd}-4{R}_{ab}{R}^{ab}+{R}^{2}$ is the second
order Lovelock (Gauss-Bonnet) Lagrangian, $\mathcal{X}_{3}$\ is
the curvature-cubed Lagrangian given by \cite{Myer1}
\begin{eqnarray}\label{X3}
\mathcal{X}_{3} &=&R_{ab}^{cd}R_{cd}^{\,\,e\,\,\,f}R_{e\,\,f}^{\,\,a\,\,\,b}+%
\frac{1}{(2n-1)(n-3)}\left(
\frac{3(3n-5)}{8}R_{abcd}R^{abcd}R\right.  \notag
\\
&&-3(n-1)R_{abcd}R^{abc}{}_{e}R^{de}+3(n+1)R_{abcd}R^{ac}R^{bd}  \notag \\
&&\left. +\,6(n-1)R_{a}{}^{b}R_{b}{}^{c}R_{c}{}^{a}-\frac{3(3n-1)}{2}%
R_{a}^{\,\,b}R_{b}^{\,\,a}R+\frac{3(n+1)}{8}R^{3}\right),
\end{eqnarray}%
and $\mathcal{X}_{4}$ is the fourth order term of
quasi-topological gravity \cite{MHD2}
\begin{eqnarray}\label{X4}
\mathcal{X}_{4}\hspace{-0.2cm} &=&\hspace{-0.2cm}c_{1}R_{abcd}R^{cdef}R_{%
\phantom{hg}{ef}%
}^{hg}R_{hg}{}^{ab}+c_{2}R_{abcd}R^{abcd}R_{ef}R^{ef}+c_{3}RR_{ab}R^{ac}R_{c}{}^{b}+c_{4}(R_{abcd}R^{abcd})^{2}
\notag \\
&&\hspace{-0.1cm}%
+c_{5}R_{ab}R^{ac}R_{cd}R^{db}+c_{6}RR_{abcd}R^{ac}R^{db}+c_{7}R_{abcd}R^{ac}R^{be}R_{%
\phantom{d}{e}}^{d}+c_{8}R_{abcd}R^{acef}R_{\phantom{b}{e}}^{b}R_{%
\phantom{d}{f}}^{d}  \notag \\
&&\hspace{-0.1cm}%
+c_{9}R_{abcd}R^{ac}R_{ef}R^{bedf}+c_{10}R^{4}+c_{11}R^{2}R_{abcd}R^{abcd}+c_{12}R^{2}R_{ab}R^{ab}
\notag \\
&&\hspace{-0.1cm}%
+c_{13}R_{abcd}R^{abef}R_{ef}{}_{g}^{c}R^{dg}+c_{14}R_{abcd}R^{aecf}R_{gehf}R^{gbhd},
\end{eqnarray}%
where the coefficients $c_{i}$ are given by \cite{MHD2}
\begin{eqnarray*}
c_{1} &=&-\left( n-1\right) \left( {n}^{7}-3\,{n}^{6}-29\,{n}^{5}+170\,{n}%
^{4}-349\,{n}^{3}+348\,{n}^{2}-180\,n+36\right) , \\
c_{2} &=&-4\,\left( n-3\right) \left( 2\,{n}^{6}-20\,{n}^{5}+65\,{n}^{4}-81\,%
{n}^{3}+13\,{n}^{2}+45\,n-18\right) , \\
c_{3} &=&-64\,\left( n-1\right) \left( 3\,{n}^{2}-8\,n+3\right) \left( {n}%
^{2}-3\,n+3\right) , \\
c_{4} &=&-{(n}^{8}-6\,{n}^{7}+12\,{n}^{6}-22\,{n}^{5}+114\,{n}^{4}-345\,{n}%
^{3}+468\,{n}^{2}-270\,n+54), \\
c_{5} &=&16\,\left( n-1\right) \left( 10\,{n}^{4}-51\,{n}^{3}+93\,{n}%
^{2}-72\,n+18\right) , \\
c_{6} &=&--32\,\left( n-1\right) ^{2}\left( n-3\right) ^{2}\left( 3\,{n}%
^{2}-8\,n+3\right) , \\
c_{7} &=&64\,\left( n-2\right) \left( n-1\right) ^{2}\left( 4\,{n}^{3}-18\,{n%
}^{2}+27\,n-9\right) , \\
c_{8} &=&-96\,\left( n-1\right) \left( n-2\right) \left( 2\,{n}^{4}-7\,{n}%
^{3}+4\,{n}^{2}+6\,n-3\right) , \\
c_{9} &=&16\left( n-1\right) ^{3}\left( 2\,{n}^{4}-26\,{n}^{3}+93\,{n}%
^{2}-117\,n+36\right) , \\
c_{10} &=&{n}^{5}-31\,{n}^{4}+168\,{n}^{3}-360\,{n}^{2}+330\,n-90, \\
c_{11} &=&2\,(6\,{n}^{6}-67\,{n}^{5}+311\,{n}^{4}-742\,{n}^{3}+936\,{n}%
^{2}-576\,n+126), \\
c_{12} &=&8\,{(}7\,{n}^{5}-47\,{n}^{4}+121\,{n}^{3}-141\,{n}^{2}+63\,n-9), \\
c_{13} &=&16\,n\left( n-1\right) \left( n-2\right) \left(
n-3\right) \left(
3\,{n}^{2}-8\,n+3\right) , \\
c_{14} &=&8\,\left( n-1\right) \left( {n}^{7}-4\,{n}^{6}-15\,{n}^{5}+122\,{n}%
^{4}-287\,{n}^{3}+297\,{n}^{2}-126\,n+18\right) .
\end{eqnarray*}%
For spherically symmetric metric, the action (\ref{Act1}) yields
second-order equations of motion in $(n+1)$-dimensional spacetimes
except in $n=2m-1$, where $m$ is the order of quasi-topological
theory \cite{MHD2}. In the remaining part of this paper, we will
show that the gravitational field equations describing by the
spacetime metric (\ref{met1}) can be recast in the form of the
first law of thermodynamics at the black hole horizon.
\section{Horizon Thermodynamics in cubic quasi-topological gravity}
In section II, we showed that the field equations of static black
hole spacetimes, with constant horizon curvature, in Einstein and
Gauss-Bonnet gravities can be reexpressed as the first law of
thermodynamics, $dE=TdS-PdV$, on the horizon. Here, we want to see
whether the above procedure works or not in other gravity theories
such as quasi-topological gravity. As we discussed, and explicitly
showed, for the Gauss-Bonnet case, one can use both the field
equations as well as the variational principle, instead of using
the field equations, and transform the equations of motion to the
first law at the spacetime horizon. However, since the field
equations for cubic \cite{Myer1} and quartic \cite{DV}
quasi-topological gravity are very long, in this section and also
the next one, we present the resulting equations from variational
principle for economic reason, which are clearly the components of
the field equations.

We shall now continue the previous procedure for the cubic term of
quasi-topological gravity. The total action of the cubic
quasi-topological gravity in $(n+1)$ dimensions can be written as
\begin{equation}\label{Act2}
I=\frac{1}{16\pi G_{n+1}}\int d^{n+1}x\sqrt{-g}[-2\Lambda +{\mu }_{1}%
\mathcal{L}_{1}+{\mu }_{2}\mathcal{L}_{2}+{\mu
}_{3}\mathcal{X}_{3}]+\int d^{n+1}x\mathcal{L}_{M}.
\end{equation}%
Varying the action (\ref{Act2}) with respect to $g^{rr}$ and
multiplying both sides in $(-g)^{-1/2}g^{rr}$, one obtains
\begin{eqnarray}\label{varr}
&&\frac{(n-1)}{16\pi
G_{n+1}}\frac{1}{Nr^{5}}\Bigg{\{}\frac{d}{dr}\left(
N^{2}f\right) \left( r^{4}+2kr^{2}\hat{\mu}_{2}l^{2}-2r^{2}\hat{\mu}%
_{2}l^{2}f+3k^{2}\hat{\mu}_{3}l^{4}-6k\hat{\mu}_{3}l^{4}f+3k\hat{\mu}%
_{3}l^{4}f^{2}\right)   \notag \\
&&+N^{2}\left[ r^{3}\left( n-2\right) \left( f-k\right) -\hat{\mu}%
_{2}l^{2}r\left( n-4\right) \left( k^{2}-2kf+f^{2}\right) -\frac{\hat{\mu}%
_{3}}{r}l^{4}\left( n-6\right) \left( -k+3k^{2}f-3kf^{2}+f^{3}\right) -\frac{%
nr^{5}}{l^{2}}\right] \Bigg{\}}=T^r_r, \nonumber \\
\end{eqnarray}%
where $T^r_r$ is given by (\ref{P}), and
\begin{equation*}
\hat{\mu}_{1}=1,\text{ \ \ }\hat{\mu}_{2}=\frac{(n-2)(n-3)}{l^{2}}\mu _{2},%
\text{ \ \ }\hat{\mu}_{3}=\frac{(n-2)(n-5)(3n^{2}-9n+4)}{8(2n-1)l^{4}}\mu
_{3}.
\end{equation*}%
Next, we evaluate Eq. (\ref{varr}) at $r=r_{+}$ and using the fact that $%
f(r_{+})=0$, to obtain
\begin{eqnarray}\label{var33}
&&\frac{N(n-1)}{16\pi G_{n+1}}\Bigg{\{}f^{\prime }(r_{+})\left( r_{+}+\frac{%
2k}{r_{+}}\hat{\mu}_{2}l^{2}+\frac{3k^{2}}{r_{+}^{3}}\hat{\mu}%
_{3}l^{4}\right) -k\left( n-2\right)   \notag   \\
&&-\frac{\hat{\mu}_{2}l^{2}}{r_{+}^{2}}k^{2}\left( n-4\right) -\frac{\hat{\mu%
}_{3}l^{4}k}{r_{+}^{4}}\left( n-6\right) -\frac{nr_{+}^{2}}{l^{2}}\Bigg{\}}%
=r_{+}^{2}P.
\end{eqnarray}%
Multiplying both sides of the above equation by the factor $\Sigma
_{k}r_{+}^{n-3}dr_{+}$, and setting $f^{\prime }(r_{+})=2\kappa $, we have
\begin{eqnarray}
&&\frac{\kappa }{2\pi }d\left[ \frac{\Sigma _{k}r_{+}^{n-1}}{4G_{n+1}}\left(
1+2k\hat{\mu}_{2}\frac{(n-1)}{(n-3)}\frac{l^{2}}{r_{+}^{2}}+3k^{2}\hat{\mu}%
_{3}\frac{(n-1)}{(n-5)}\frac{l^{4}}{r_{+}^{4}}\right) \right]   \notag \\
&&-d\left[ \frac{(n-1)\Sigma _{k}r_{+}^{n-2}}{16\pi G_{n+1}}\left( k+\frac{%
k^{2}\hat{\mu}_{2}l^{2}}{r_{+}^{2}}+\frac{k^{3}\hat{\mu}_{3}l^{4}}{r_{+}^{4}}%
+\frac{r_{+}^{2}}{l^{2}}\right) \right] =P\Sigma _{k}r{+}^{n-1}dr_{+}.
\end{eqnarray}%
We can rewrite this equation in the form,
\begin{eqnarray}\label{flt22}
&&Td\left[ \frac{\Sigma _{k}r_{+}^{n-1}}{4G_{n+1}}\left( 1+2k\hat{\mu}_{2}%
\frac{(n-1)}{(n-3)}\frac{l^{2}}{r_{+}^{2}}+3k^{2}\hat{\mu}_{3}\frac{(n-1)}{%
(n-5)}\frac{l^{4}}{r_{+}^{4}}\right) \right]   \notag \\
&&-d\left[ \frac{(n-1)\Sigma _{k}r_{+}^{n}}{16\pi G_{n+1}l^{2}}\left( 1+k%
\frac{l^{2}}{r_{+}^{2}}+k^{2}\hat{\mu}_{2}\frac{l^{4}}{r_{+}^{4}}+k^{3}\hat{%
\mu}_{3}\frac{l^{6}}{r_{+}^{6}}\right) \right] =PdV.
\end{eqnarray}%
The first term in the left hand side of the above equation is in the form $%
TdS$, and so one may recognize the entropy expression for the horizon in
cubic quasi-topological gravity as,
\begin{equation}\label{S3}
S=\frac{\Sigma _{k}r_{+}^{n-1}}{4G_{n+1}}\left( 1+2k\hat{\mu}_{2}\frac{(n-1)%
}{(n-3)}\frac{l^{2}}{r_{+}^{2}}+3k^{2}\hat{\mu}_{3}\frac{(n-1)}{(n-5)}\frac{%
l^{4}}{r_{+}^{4}}\right).
\end{equation}%
According to the first low of thermodynamics, we can interpret the second
term in the left hand side of (\ref{flt22}) as the energy of the system,
\begin{equation}\label{E3}
E=\frac{(n-1)\Sigma _{k}r_{+}^{n}}{16\pi G_{n+1}l^{2}}\left( 1+k\frac{l^{2}}{%
r_{+}^{2}}+k^{2}\hat{\mu}_{2}\frac{l^{4}}{r_{+}^{4}}+k^{3}\hat{\mu}_{3}\frac{%
l^{6}}{r_{+}^{6}}\right) .
\end{equation}%
These expressions for entropy and energy of the black holes
coincide exactly with those obtained in quasi-topological gravity
by solving the field equations and using the Wald's method
\cite{Myer1}. Here we arrived at the same result by transforming
the field equations to the form of the first law on the black hole
horizon. This indicates that the approach presented here is enough
powerful and further reveals the deep connection between the
gravitational field equations and the first law of thermodynamics
on the horizon of the black hole. Again, one can show that
although $P$ and $\rho $ are different in general but they are the
same at the horizon.
\section{Horizon Thermodynamics of Quasi-topological Gravity}
In this section, we would like to extend the above study to the case of $m$%
-th order quasi-topological gravity. First, we consider the
quartic quasi-topological gravity. Varying the action (\ref{Act1})
with respect to $g^{rr}$ and multiplying both sides in
$(-g)^{-1/2}g^{rr}$, we obtain
\begin{eqnarray}\label{Var4}
&&\frac{(n-1)}{16\pi
G_{n+1}}\frac{1}{Nr^{7}}\Bigg{\{}\frac{d}{dr}\left(
N^{2}f\right) \left( r^{6}+2kr^{4}\hat{\mu}_{2}l^{2}-2r^{4}\hat{\mu}%
_{2}l^{2}f+3k^{2}r^{2}\hat{\mu}_{3}l^{4}-6kr^{2}\hat{\mu}_{3}l^{4}f+3r^{2}%
\hat{\mu}_{3}l^{4}f^{2}\right)   \notag \\
&&+N^{2}[f^{\prime }\left( 4k\hat{\mu}_{4}l^{6}-12k^{2}\hat{\mu}%
_{4}l^{6}f+12k\hat{\mu}_{4}l^{6}f^{2}-4\hat{\mu}_{4}l^{6}f^{3}\right)
+r^{5}\left( n-2\right) \left( f-k\right) +\hat{\mu}_{2}l^{2}r^{3}\left(
n-4\right) \left( 2kf-k^{2}-f^{2}\right)  \nonumber\\
&& -\hat{\mu}_{3}l^{4}r\left( n-6\right) \left( k-3k^{2}f+3kf^{2}-f^{3}\right) +%
\frac{\hat{\mu}_{4}}{r}l^{6}\left( n-8\right) \left(
-k^{2}+4kf-6k^{2}f^{2}+4kf^{3}-f^{4}\right)
-\frac{nr^{7}}{l^{2}}]\Bigg{\}} =T_{r}^{r},
\end{eqnarray}%
where
\begin{equation*}
\hat{\mu}_{4}=\frac{%
n(n-1)(n-2)^{2}(n-3)(n-7)(n^{5}-15n^{4}+72n^{3}-156n^{2}+150n-42)}{l^{6}}\mu
_{4}.
\end{equation*}%
Evaluating Eq. (\ref{Var4}) at $r=r_{+}$ and setting $f(r_{+})=0$, we
arrive at
\begin{eqnarray}\label{fltq4}
&&\frac{N(n-1)}{16\pi G_{n+1}}\Bigg{\{}f^{\prime }\left( r_{+}+\frac{2k}{%
r_{+}}\hat{\mu}_{2}l^{2}+\frac{3k^{2}}{r_{+}^{3}}\hat{\mu}_{3}l^{4}+\frac{4k%
}{r_{+}^{5}}\hat{\mu}_{4}l^{6}\right) -k\left( n-2\right)   \notag \\
&&-\hat{\mu}_{2}\frac{k^{2}l^{2}}{r_{+}^{2}}\left( n-4\right) -\hat{\mu}_{3}%
\frac{kl^{4}}{r_{+}^{4}}\left( n-6\right) -\frac{\hat{\mu}_{4}l^{6}k^{2}}{%
r_{+}^{6}}\left( n-8\right)
-\frac{nr_{+}^{2}}{l^{2}}\Bigg{\}}=Pr_{+}^{2},
\end{eqnarray}%
Multiplying both sides of the above equation by the factor $\Sigma
_{k}r_{+}^{n-3}dr_{+}$ and setting $f^{\prime }(r_{+})=2\kappa $, we get
\begin{eqnarray}\label{fltq5}
&&\frac{\kappa }{2\pi }d\left[ \frac{\Sigma
_{k}r_{+}^{n-1}}{4G_{n+1}}\left(
1+2k\hat{\mu}_{2}\frac{(n-1)}{(n-3)}\frac{l^{2}}{r_{+}^{2}}+3k^{2}\hat{\mu}%
_{3}\frac{(n-1)}{(n-5)}\frac{l^{4}}{r_{+}^{4}}+4k^{3}\hat{\mu}_{4}\frac{(n-1)%
}{(n-7)}\frac{l^{6}}{r_{+}^{6}}\right) \right]   \notag \\
&&-d\left[ \frac{(n-1)\Sigma _{k}r_{+}^{n-2}}{16\pi G_{n+1}}\left( k+\frac{%
\hat{\mu}_{2}l^{2}k^{2}}{r_{+}^{2}}+\frac{\hat{\mu}_{3}l^{4}k^{3}}{r_{+}^{4}}%
+\frac{\hat{\mu}_{4}l^{6}k^{4}}{r_{+}^{6}}+\frac{r_{+}^{2}}{l^{2}}\right) %
\right] =P\Sigma _{k}r{+}^{n-1}dr_{+}
\end{eqnarray}%
Using the definition $T=\kappa /2\pi $, the above equation can be rewritten
in the form
\begin{eqnarray}\label{fltq6}
&&Td\left[ \frac{\Sigma _{k}r_{+}^{n-1}}{4G_{n+1}}\left( 1+2k\hat{\mu}_{2}%
\frac{(n-1)}{(n-3)}\frac{l^{2}}{r_{+}^{2}}+3k^{2}\hat{\mu}_{3}\frac{(n-1)}{%
(n-5)}\frac{l^{4}}{r_{+}^{4}}+4k^{3}\hat{\mu}_{4}\frac{(n-1)}{(n-7)}\frac{%
l^{6}}{r_{+}^{6}}\right) \right]   \notag \\
&&-d\left[ \frac{(n-1)\Sigma _{k}r_{+}^{n}}{16\pi G_{n+1}l^{2}}\left( 1+k%
\frac{l^{2}}{r_{+}^{2}}+\hat{\mu}_{2}k^{2}\frac{l^{4}}{r_{+}^{4}}+\hat{\mu}%
_{3}k^{3}\frac{l^{6}}{r_{+}^{6}}+\hat{\mu}_{4}k^{4}\frac{l^{8}}{r_{+}^{8}}%
\right) \right] =PdV.
\end{eqnarray}%
Equation (\ref{fltq6}) is nothing, but the first law of
thermodynamics on the horizon, $TdS-dE=PdV$. We can define the
entropy expression as
\begin{equation}\label{SS1}
S=\frac{\Sigma _{k}r_{+}^{n-1}}{4G_{n+1}}\left( 1+2k\hat{\mu}_{2}\frac{(n-1)%
}{(n-3)}\frac{l^{2}}{r_{+}^{2}}+3k^{2}\hat{\mu}_{3}\frac{(n-1)}{(n-5)}\frac{%
l^{4}}{r_{+}^{4}}+4k^{3}\hat{\mu}_{4}\frac{(n-1)}{(n-7)}\frac{l^{6}}{%
r_{+}^{6}}\right) ,
\end{equation}%
and the total energy (mass) of the black hole as
\begin{equation}\label{E4th}
E=\frac{(n-1)\Sigma _{k}r_{+}^{n}}{16\pi G_{n+1}l^{2}}\left( 1+k\frac{l^{2}}{%
r_{+}^{2}}+\hat{\mu}_{2}k^{2}\frac{l^{4}}{r_{+}^{4}}+\hat{\mu}_{3}k^{3}\frac{%
l^{6}}{r_{+}^{6}}+\hat{\mu}_{4}k^{4}\frac{l^{8}}{r_{+}^{8}}\right)
.
\end{equation}%
The obtained expressions for entropy and energy are precisely the
expressions calculated by other authors for topological black holes in
quartic quasi-topological gravity \cite{MHD2}. In this way we show that the
field equations of quartic quasi-topological gravity can be transformed to
the form of the first law of thermodynamics on the event horizon of $(n+1)$%
-dimensional topological black holes.

Having the results for the cubic and quartic cases at hand, one may
conjecture that there exists similar connection for $m$-th order
quasi-topological gravity in $n\neq 2m-1$ dimensions. For the topological
black holes with metric (\ref{met1}), one may conjecture that the gravity
part of the action (\ref{Act1}) is given by \cite{MHD2}
\begin{equation}\label{ActKol}
I_{G}=\int dtdrN(r)\left( r^{n}\sum_{i=0}^{m}\hat{\mu}_{i}\left[
l^{2}r^{-2}(k-f)\right] ^{i}\right) ^{\prime },
\end{equation}%
where $\hat{\mu}_{i}$ are coefficients of the $i$-th powered curvature term
with $\hat{\mu}_{1}=1$. The corresponding field equations evaluated on the
horizon may be rewritten as
\begin{equation}\label{Varmor}
Td\left[ \frac{\Sigma _{k}r_{+}^{n-1}}{4G_{n+1}}\sum_{i=1}^{m}i\frac{(n-1)}{%
(n+1-2i)}\frac{\hat{\mu}_{i}k^{i-1}l^{2i-2}}{r_{+}^{2i-2}}\right] -d\left[
\frac{(n-1)\Sigma _{k}r_{+}^{n}}{16\pi G_{n+1}l^{2}}\left( 1+\sum_{i=1}^{m}%
\hat{\mu}_{i}k^{i}\frac{l^{2i}}{r_{+}^{2i}}\right) \right] =PdV.
\end{equation}
Equation (\ref{Varmor}) is the first law of thermodynamics provided we define
\begin{eqnarray}\label{Sg}
S &=&\frac{\Sigma _{k}r_{+}^{n-1}}{4G_{n+1}}\sum_{i=1}^{m}i\frac{(n-1)}{%
(n+1-2i)}\frac{\hat{\mu}_{i}k^{i-1}l^{2i-2}}{r_{+}^{2i-2}},  \label{Eg} \\
E &=&\frac{(n-1)\Sigma _{k}r_{+}^{n}}{16\pi G_{n+1}l^{2}}\left(
1+\sum_{i=1}^{m}\hat{\mu}_{i}k^{i}\frac{l^{2i}}{r_{+}^{2i}}\right)
.
\end{eqnarray}%
These are the most general expressions for entropy and energy of static
black hole spacetimes with spherical, hyperbolic or planar horizon topology
in the most general quasi-topological theory of gravity. We expect to
confirm our general results (\ref{Eg}) and (\ref{Sg}) in the future by
solving explicitly the field equations. Again, one can show that although $P$
and $\rho $ are different in general but they are the same at the horizon.
\section{CONCLUSIONS\label{con}}
According to the black hole thermodynamics, a black hole can be
regarded as a thermodynamic system which has entropy and
temperature associated with its horizon. Since the discovery of
black hole thermodynamics in $1970$'s physicists have been
speculating that there should be some deep connection between
thermodynamics and gravity. This is due to the fact that
thermodynamic quantities of black holes such as temperature and
entropy are closely related to their geometrical quantities such
as surface gravity and horizon area. In this paper, we have
investigated the thermodynamics of topological black holes, with
spherical, hyperbolic or planar horizon topology, in
quasi-topological theory of gravity. We showed that one can always
rewrite the field equations of quasi-topological gravity in the
form of the first law of thermodynamics, $dE=TdS-PdV$, at the
black hole horizon. This procedure allows us to obtain the entropy
and the mass expressions in terms of the radius of black hole
horizon, which are exactly the same as those resulting from the
Wald's method for black hole entropy and the Hamiltonian approach
for black hole mass. The novelty and advantages of the present
study is that in the process of deriving the entropy and the mass
of black holes, we have not solved the field equation of the
quasi-topological theory and we had no difficulties of Wald's
method for calculating entropy. This completely differs from the
previous works in the literature \cite{Myer1,MHD2}.

The thermodynamic interpretation of the gravitational field
equations in the most general quasi-topological gravity indicates
that the connection between thermodynamics and gravity is not just
an accident, but something with deep physical meaning. On the
other hand the disclosed relation on the field equations and the
first law of thermodynamics on the black hole horizon also sheds
the light on holography, since the gravitational field equations
persists the information in the bulk and the first law of
thermodynamics on the event horizon contains the information on
the boundary. Our study shows that the approach here is powerful
to find an expression of entropy in term of the horizon radius. It
could help to extract an expression of entropy associated with the
event horizon in quasi-topological gravity, which is useful in
studying the thermodynamical properties of black holes in this
theory.
\acknowledgments{We thank from the Research Council of Shiraz
University. This work has been supported financially by Research
Institute for Astronomy \& Astrophysics of Maragha (RIAAM), Iran.

\end{document}